\UseRawInputEncoding

\documentclass[12pt]{iopart}
\usepackage{iopams}  
\usepackage{amssymb}
\usepackage{bm}
\usepackage{epsfig}
\usepackage{wasysym}

\usepackage{color}

\begin{document}

\title{Optical absorption measurement of spin Berry curvature and spin Chern marker}


\author{Wei Chen$^{1}$}



\address{$^{1}$Department of Physics, PUC-Rio, Rio de Janeiro 22451-900, Brazil}

\ead{wchen@puc-rio.br}

\begin{abstract}

In two-dimensional time-reversal symmetric topological insulators described by Dirac models, the ${\mathbb Z}_{2}$ topological invariant can be described by the spin Chern number. We present a linear response theory for the spin Berry curvature that integrates to the spin Chern number, and introduce its spectral function that can be measured at finite temperature by momentum- and spin-resolved circular dichroism, which may be achieved by pump-probe type of experiments using spin- and time-resolved ARPES. As a result, the sign of the Pfaffian of the ${\mathbb Z}_{2}$ invariant can be directly measured. The spin Chern number expressed in real space yields a spin Chern marker, whose spatial variation may be measured by circular dichroism and spin-resolved photoemission with a spatial resolution. A spin Chern correlator and a nonlocal spin Chern marker are further proposed to characterize the quantum criticality near topological phase transitions, which are shown to take the form of overlaps between spin-selected Wannier states.

\end{abstract}

%
%
%
%
%

\section{Introduction}

The topological order has been an intriguing subject over the past few decades since the discovery of quantum Hall effect\cite{vonKlitzing80}. Particularly in two-dimensional (2D) time-reversal (TR) breaking materials, the topological order manifests as the quantum anomalous Hall effect\cite{Haldane88}, in which the quantized Hall conductance is calculated from the momentum-integration of the Berry curvature of the valence band Bloch states\cite{Thouless82}. On the other hand, for 2D TR-symmetric topological insulators (TIs) that belong to class AII on the Altland-Zirnbauer symmetry classification table\cite{Schnyder08,Ryu10,Kitaev09,Chiu16}, the topological order is described by a binary ${\mathbb Z}_{2}$ invariant\cite{Kane05,Fu07}. The defining feature of these materials is the quantum spin Hall effect (QSHE)\cite{Kane05,Kane05_2,Bernevig06} that has been observed experimentally in heterostructures\cite{Konig07,Liu08,Knez11}, bilayer\cite{Yang12,Drozdov14}, and monolayer materials\cite{Tang17,Jia17,Peng17,Reis17,Fei17,Wu18,Deng18,Xu18,Zhu19,Shi19}, in which propagating spinful edge states occurs when the system enters topologically nontrivial phase, often linked to an equilibrium edge spin current\cite{Chen20_absence_edge_current}. In addition, away from the edge, although the bulk of the material seems to be indistinguishable from a featureless insulator, it has been proposed that a locally defined spin Bott index that measures the commutativity of the projected position
operators can correctly capture the ${\mathbb Z}_{2}$ invariant\cite{Huang18,Huang18_2}. However, the possibility of measuring the ${\mathbb Z}_{2}$ invariant deep inside the bulk without involving the edge state seems to remain elusive at present.


In this paper, we elaborate that optical absorption measurements on the bulk bands of 2D TIs, without involving any properties of the edge states, can be used to identify the ${\mathbb Z}_{2}$ invariant. Focusing on the ideal 2D TIs that have no other complications like spin-orbit coupling, we elaborate that the ${\mathbb Z}_{2}$ invariant can be measured in the bulk either by momentum space or real space optical absorption measurements. In this ideal situation, the ${\mathbb Z}_{2}$ invariant is given by the spin Chern number calculated from the momentum-integration of the spin Berry curvature\cite{Yao05,Sheng06,Prodan07,Fukui07,Gradhand12,Sinova15,Zhang17_spin_Berry,Ulcakar18}, defined as the Berry curvature of the spin up electrons in the valence bands minus that of the spin down electrons. We first connect this problem to a recently developed linear response theory, which shows that Berry curvature can be measured by the momentum-resolved exciton absorption rate in circular dichroism experiments, a method that is applicable to real materials at finite temperature that contain complications like many-body interactions and disorder\cite{Chen22_dressed_Berry_metric}. Through adding spin degrees of freedom, we demonstrate that the spin Berry curvature can also be measured in the same manner by momentum-resolved exciton absorption rate, provided the circular dichroism experiment has a spin resolution. Because the ${\mathbb Z}_{2}$ invariant in this ideal case is entirely determined by the signs of the spin Berry curvature at the high symmetry points (HSPs) of the Brillouin zone (BZ), this method can efficiently determine the signs by simply measuring which spin species has a larger circular dichroism at these HSPs.

We proceed to propose a real space measurement for the ${\mathbb Z}_{2}$ invariant through the spin Chern marker\cite{Prodan10_2,Prodan11}, which is a real space topological invariant constructed from writing the spin Chern number into spin-selected projectors to filled and empty states of lattice models\cite{Prodan10,Prodan10_2,Prodan11,Bianco11,Bianco13,Marrazzo17}. Through adopting a linear response theory developed recently for the spinless version of the theory, the spin Chern marker is shown to be measurable by real space circular dichroism experiments combined with spin- and time-resolved photoemission, which may be used to image how the ${\mathbb Z}_{2}$ invariant is influenced by various sources of inhomogeneity in real space, such as impurities or grain boundaries. Furthermore, a spin Chern correlator and a nonlocal spin Chern marker are proposed to detect the quantum criticality near topological phase transitions (TPTs), and their relation to Wannier states and nonlocal susceptibilities will be elaborated.

\section{Measurement in momentum space}

\subsection{Spin Berry curvature as the Jacobian of the wrapping number \label{sec:spin_Berry_Jacobian}}

To formulate a theory for 2D class AII materials that can connect the ${\mathbb Z}_{2}$ invariant, projectors to eigenstates, Berry curvature, and Wannier states all at once, our strategy is to resort to a recently proposed universal topological invariant that is applicable to any dimension and symmetry class on the Altland-Zirnbauer symmetry classification table, provided the system is described by a Dirac model\cite{vonGersdorff21_unification}. This requirement restricts our discussion to 2D class AII materials whose spin polarization at any momentum is the same, i.e., no spin-orbit coupling. To draw relevance to known materials, we will discuss the problem using the prototype Bernevig-Hughes-Zhang (BHZ) model\cite{Bernevig06,Konig07}, which composes of spinful $s$ and $p$ orbitals $\psi=\left(s\uparrow,p\uparrow,s\downarrow,p\downarrow\right)^{T}$, and uses the representation for the Dirac matrices
\begin{eqnarray}
\left\{\gamma_{0},\gamma_{1},\gamma_{2},\gamma_{3},\gamma_{4}\right\}
=\left\{I\otimes s^{z},\sigma^{z}\otimes s^{x},I\otimes s^{y},\sigma^{x}\otimes s^{x},\sigma^{y}\otimes s^{x}\right\}.
\end{eqnarray}
In momentum space, the single-particle Hamiltonian reads\cite{Bernevig13} 
\begin{eqnarray}
&&H({\bf k})=\sum_{i=0}^{2}d_{i}({\bf k})\gamma_{i}
=A\sin k_{x}\gamma_{1}+A\sin k_{y}\gamma_{2}
\nonumber \\
&&+\left(M-4B+2B\cos k_{x}+2B\cos k_{y}\right)\gamma_{0}
\nonumber \\
&&=\left(
\begin{array}{cc}
h({\bf k}) & 0 \\
0 & h^{\ast}(-{\bf k})
\end{array}
\right),
\label{BHZ_Dirac_Hamiltonian}
\end{eqnarray}
where $h({\bf k})=\sum_{i=1}^{3}d_{i}({\bf k})\sigma^{i}$, $A$ and $B$ are kinetic parameters, and $M$ is the mass term that controls the topological order. The unit vector $n_{i}=d_{i}/d$ defines a spectrally flattened Hamiltonian by $\tilde{Q}={\bf n}({\bf k})\cdot{\boldsymbol\gamma}$.

The universal topological invariant counts the number of times the BZ torus $T^{D}$ wraps around the target sphere $S^{D}$ formed by the ${\bf n}({\bf k})$, which is referred to as the wrapping number or degree of the map ${\rm deg}[{\bf n}]$, and take the following cyclic derivative form in $D$ spatial dimensions\cite{vonGersdorff21_unification} 
\begin{eqnarray}
{\rm deg}[{\bf n}]&=&\frac{1}{V_{D}}\int d^{D}k\,\varepsilon^{i_{0}i_{1}...i_{D}}\frac{1}{d^{D+1}}d_{i_{0}}\partial_{1}d_{i_{1}}...\partial_{D}d_{i_{D}}
\nonumber \\
&=&\frac{1}{V_{D}}\int d^{D}k\,\varepsilon^{i_{0}i_{1}...i_{D}}n_{i_{0}}\partial_{1}n_{i_{1}}...\partial_{D}n_{i_{D}},
\label{wrapping_number}
\end{eqnarray}
where $V_{D}=2\pi^{(D+1)/2}/\Gamma((D+1)/2)$ is the volume of the $D$-sphere. The integrand $\varepsilon^{i_{0}i_{1}...i_{D}}n_{i_{0}}\partial_{1}n_{i_{1}}...\partial_{D}n_{i_{D}}$ represents the Jacobian of the map $T^{D}\rightarrow S^{D}$. For our 2D class AII problem at hand, we have $D=2$, and the ${\mathbb Z}_{2}$ topological invariant is given by $(-1)^{{\rm deg}[{\bf n}]}$. Our aim is to express the Jacobian in terms of the spectrally flattened Hamiltonian by the form $n_{i}\partial_{x}n_{j}\partial_{y}n_{k}={\rm Tr}\left[W\tilde{Q}(d\tilde{Q})^{2}\right]/N$ such that it can further be written into some product of conduction and valence band states, where $W$ is a matrix to be determined, $N$ is a normalization factor, and $(d\tilde{Q})^{2}=\partial_{x}\tilde{Q}\partial_{y}\tilde{Q}-\partial_{y}\tilde{Q}\partial_{x}\tilde{Q}$.

For the BHZ model in Eq.~(\ref{BHZ_Dirac_Hamiltonian}), the spectrally flattened Hamiltonian takes the form
\begin{eqnarray}
\tilde{Q}=\left(\begin{array}{cccc}
n_{0} & n_{-} & & \\
n_{+} & -n_{0} & & \\
 & & n_{0} & -n_{+} \\
 & & -n_{-} & -n_{0}
\end{array}\right),
\end{eqnarray}
where $n_{\pm}=n_{1}\pm in_{2}$. It is straight forward to show that 
\begin{eqnarray}
{\rm Tr}\left[\gamma_{3}\gamma_{4}\tilde{Q}(d\tilde{Q})^{2}\right]
=8i\varepsilon^{ijk}n_{i}\partial_{x}n_{j}\partial_{y}n_{k},
\label{2DclassAII_TrQdQ_ndn}
\end{eqnarray}
with $n_{i}=\left\{n_{0},n_{1},n_{2}\right\}$. The motivation of inserting $W=\gamma_{3}\gamma_{4}$ comes from the fact that $\gamma_{3}$ and $\gamma_{4}$ are the Dirac matrices that have been omitted in the Dirac Hamiltonian. In order to remove the matrix product in ${\rm Tr}\left[W\tilde{Q}(d\tilde{Q})^{2}\right]$, we must insert $W=\gamma_{3}\gamma_{4}$ such that ${\rm Tr}\left[\gamma_{0}\gamma_{1}\gamma_{2}\gamma_{3}\gamma_{4}\right]$ is only a constant, leaving only the cyclic derivative $\varepsilon^{ijk}n_{i}\partial_{x}n_{j}\partial_{y}n_{k}$ that is proportional to the integrand in Eq.~(\ref{wrapping_number}).

Interestingly, the inserted matrix is simply the spin operator $W=\gamma_{3}\gamma_{4}=i\sigma^{z}\otimes I$. Writing the spectrally flattened Hamiltonian into the projection operators 
\begin{eqnarray}
\tilde{Q}=Q-P=\sum_{m}|m\rangle\langle m|-\sum_{n}|n\rangle\langle n|,
\end{eqnarray}
where $|n\rangle$ denotes the valence band and $|m\rangle$ the conduction band eigenstates. Applying $\gamma_{3}\gamma_{4}$ on these states simply gets the spin polarization of the states $\gamma_{3}\gamma_{4}|n\rangle=i\sigma_{n}^{z}|n\rangle$ and $\gamma_{3}\gamma_{4}|m\rangle=i\sigma_{m}^{z}|m\rangle$ due to the block-diagonal form of Eq.~(\ref{BHZ_Dirac_Hamiltonian}), where $\sigma_{n}^{z}=\pm 1$ is the spin polarization. As a result, Eq.~(\ref{2DclassAII_TrQdQ_ndn}) expressed in terms of the projectors is
\begin{eqnarray}
&&{\rm Tr}\left[\gamma_{3}\gamma_{4}\tilde{Q}(d\tilde{Q})^{2}\right]
=-8i\sum_{nm}\left\{\langle\partial_{x}n|\sigma_{m}^{z}|m\rangle\langle m|\partial_{y}n\rangle-\langle\partial_{y}n|\sigma_{m}^{z}|m\rangle\langle m|\partial_{x}n\rangle\right\}
\nonumber \\
&&=-8i\sum_{nm}\left\{\langle\partial_{x}n|{\hat \sigma}^{z}|\partial_{y}n\rangle-\langle\partial_{y}n|{\hat \sigma}^{z}|\partial_{x}n\rangle\right\}
\nonumber \\
&&=-8\left(\Omega_{xy}^{\uparrow}-\Omega_{xy}^{\downarrow}\right)
\equiv-16\,\Omega_{xy}^{s}.
\label{2DclassAII_TrQdQ_PxQy}
\end{eqnarray}
We have just proved that the Jacobian of the map from $T^{2}\rightarrow S^{2}$ for 2D class AII is equivalently the spin Berry curvature $\Omega_{xy}^{s}=(\Omega_{xy}^{\uparrow}-\Omega_{xy}^{\downarrow})/2$\cite{Yao05,Sheng06,Fukui07,Prodan07,Gradhand12,Sinova15,Zhang17_spin_Berry,Ulcakar18}.

Conventionally, the ${\mathbb Z}_{2}$ invariant is calculated from the Pfaffian of the TR operator ${\rm Pf}\left[\langle n({\bf k}_{0})|T|n'({\bf k}_{0})\rangle\right]$ at HSPs ${\bf k}_{0}$\cite{Kane05,Fu07}. The product of the signs of the Pfaffian at all the HSPs give the ${\mathbb Z}_{2}$ invariant $\nu$ by $(-1)^{\nu}=\prod_{\bf k_{0}}{\rm Sgn}\left\{{\rm Pf}\left[\langle n({\bf k}_{0})|T|n'({\bf k}_{0})\rangle\right]\right\}$. In our ideal Dirac model described by Eq.~(\ref{BHZ_Dirac_Hamiltonian}), the sign of Pfaffian at a ${\bf k}_{0}$ is equal to the sign of the spin Berry curvature at the same ${\bf k}_{0}$
\begin{eqnarray}
{\rm Sgn}\left\{{\rm Pf}\left[\langle n({\bf k}_{0})|T|n'({\bf k}_{0})\rangle\right]\right\}={\rm Sgn}\left[\Omega^{s}_{xy}({\bf k}_{0})\right]={\rm Sgn}\left[d_{3}\right].
\label{sign_Pfaffian_sign_spin_Berry}
\end{eqnarray}
This coincidence suggests that at least in this ideal case, direct measurement of the signs of the spin Berry curvature at HSPs can determine the ${\mathbb Z}_{2}$ invariant. We should advice such an experiment in the following section.



\subsection{Linear response theory and measurement of spin Berry curvature \label{sec:linear_response_kspace}}

We now present a linear response theory that links the spin Berry curvature to experimental measurables. The theory is completely analogous to that of the Berry curvature proposed recently\cite{Chen22_dressed_Berry_metric}. Consider the application of a polarized external electric field $E^{\mu}$ that couples to the operator $i\partial_\mu$ via the dipole energy\cite{Karplus54,Ozawa18,vonGersdorff21_metric_curvature} 
\begin{eqnarray}
\delta h({\bf k})=-iqE^{\mu}\partial_{\mu},
\label{deltah_Efield}
\end{eqnarray}
where $q$ is the electron charge. In the second-quantization form, the perturbation reads 
\begin{eqnarray}
&&\delta H({\bf k})=\sum_{nn'\sigma}\langle n{\bf k}\sigma|\delta h({\bf k})|n'{\bf k}\sigma\rangle \,c_{n{\bf k}\sigma}^{\dag}c_{n'{\bf k}\sigma}=-qE^{\mu}\sum_{\sigma}U_{\mu}^{\sigma}({\bf k}),
\end{eqnarray}
where we have written explicitly the spin eigenvalue $\sigma$ originated from the block-diagonal Hamiltonian in Eq.~(\ref{BHZ_Dirac_Hamiltonian}) and the momentum dependence of the eigenstates, and used the fact that the perturbation in Eq.~(\ref{deltah_Efield}) does not mix up the spin species so $\langle n{\bf k}\sigma|\delta h({\bf k})|n'{\bf k}\sigma '\rangle=0$ if $\sigma\neq\sigma '$. The charge polarization operator $U_{\mu}^{\sigma}$ of electrons with spin $\sigma$ is defined by 
\begin{eqnarray}
&&U_{\mu}^{\sigma}({\bf k})=\sum_{nn'}{\cal A}_{\mu}^{\sigma nn'}({\bf k})\,c_{n{\bf k}\sigma}^{\dag}c_{n'{\bf k}\sigma}=-U^{\sigma\dag}_{\mu}({\bf k}),
\nonumber \\
&&{\cal A}_{\mu}^{\sigma nn'}({\bf k})=\langle n{\bf k}\sigma|i\partial_{\mu}|n'{\bf k}\sigma\rangle\equiv {\cal A}_{\mu}^{\sigma nn'}.
\end{eqnarray}
where ${\cal A}_{\mu}^{\sigma nn'}$ is the non-Abelian gauge field defined from the eigenstates with the same spin $\sigma$.

Our aim is to calculate the susceptibility $\chi_{\mu\nu}^{\sigma}$ of the ensemble average of the charge polarization operator $U_{\mu}^{\sigma}$ of spin $\sigma$ 
\begin{eqnarray}
\langle U_{\mu}^{\sigma}({\bf k},t)\rangle=\chi_{\mu\nu}^{\sigma}({\bf k},t)qE^{\nu}(t),
\end{eqnarray}
caused by the oscillating electric field $E^{\nu}(t)=E^{\nu}e^{-i\omega t}$, where the operators are defined in the Heisenberg picture $U_{\mu}({\bf k},t)=e^{i(H_{0}+H')t}U_{\mu}({\bf k})e^{-i(H_{0}+H')t}$, with $H_{0}$ and $H'$ the bare and the interacting Hamiltonians, respectively. The retarded susceptibility may be calculated conveniently using Matsubara frequency and then performing an analytical continuation  $i\omega\rightarrow\omega+i\eta$. The real part of the antisymmetrized retarded susceptibility has been proposed as the Berry curvature spectral function\cite{Chen22_dressed_Berry_metric}, which now carries a spin index $\sigma$
\begin{eqnarray}
&&\Omega^{d\sigma}_{xy}({\bf k},\omega)\equiv-\frac{1}{\pi}{\rm Re}\left[\chi_{xy}^{\sigma}({\bf k},\omega)-\chi_{yx}^{\sigma}({\bf k},\omega)\right],
\end{eqnarray}
where the superscript $d$ stands for "dressed", since the formalism can incorporate many-body interactions. The dressed Berry curvature is calculated from the integration of the spectral functions over positive frequency
\begin{eqnarray}
&&\Omega^{d\sigma}_{xy}({\bf k})=\int_{0}^{\infty}d\omega\,\Omega^{d\sigma}_{xy}({\bf k},\omega).
\label{gmunu_frequency_integration}
\end{eqnarray}
It can be easily verified that at the zero temperature $T=0$ and noninteracting $H'=0$ limit, the dressed spin Berry curvature $\Omega^{ds}_{xy}\equiv (\Omega^{d\uparrow}_{xy}-\Omega^{d\downarrow}_{xy})/2$ recovers the spin Berry curvature $\Omega^{s}_{xy}$ in Eq.~(\ref{2DclassAII_TrQdQ_PxQy}).


\begin{figure}[ht]
\begin{center}
\includegraphics[clip=true,width=0.8\columnwidth]{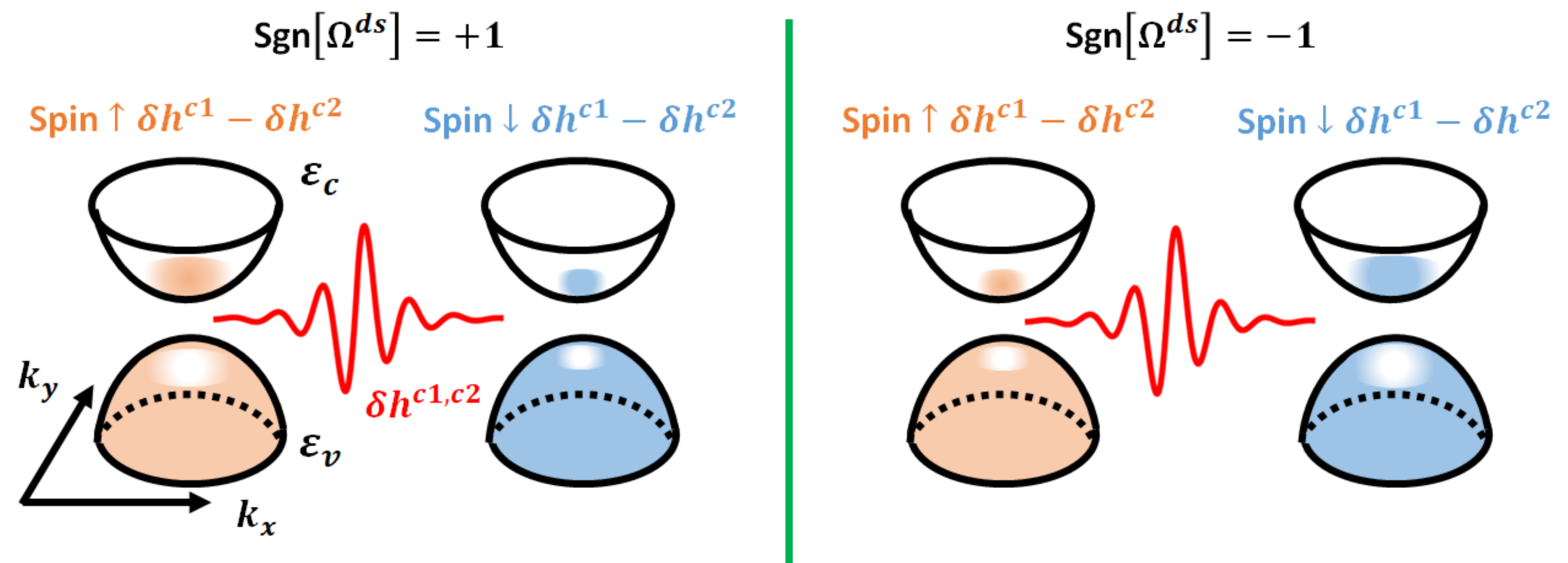}
\caption{The proposed experiment to measure the sign of dressed spin Berry curvature ${\rm Sgn}[\Omega^{ds}({\bf k})]$ at an HSP ${\bf k}_{0}$. In the spin-resolved circular dichroism experiment, if more spin up electrons are excited from the valence $\varepsilon_{v}$ to the conduction band $\varepsilon_{c}$ than spin down, then ${\rm Sgn}[\Omega^{ds}]=+1$, otherwise ${\rm Sgn}[\Omega^{ds}]=-1$. The orange and blue colors indicate the spin density of the two bands in the experiment, which may be measured by spin- and time-resolved ARPES in the pump-probe type of experiments. } 
\label{fig:circ_dic_spin_Berry}
\end{center}
\end{figure}

\subsection{Measurement of spin Berry curvature}

The linear response theory implies the a circular dichroism measurement protocol for the spin Berry curvature, as we shall see below. In fact, the question of whether the Pfaffian in Eq.~(\ref{sign_Pfaffian_sign_spin_Berry}) can be measured locally by light has been discussed recently\cite{LeHur22}. Consider the application of electric field in the two spatial directions and with a phase difference $\pm i$\cite{Repellin19}
\begin{eqnarray}
\delta h^{c1,c2}=\sum_{\sigma}\left(U_{x}^{\sigma\dag}\pm iU_{y}^{\sigma\dag}\right)qE_{0}e^{-i\omega t},
\label{hc1c2_kspace}
\end{eqnarray}
which correspond to the two circular polarizations. The field
causes the charge polarization in each spin species by
\begin{eqnarray}
&&\langle U_{x}^{\sigma}({\bf k},t)\mp iU_{y}^{\sigma}({\bf k},t)\rangle=\chi_{\sigma}^{c1,c2}({\bf k},t)qE_{0}e^{-i\omega t},
\end{eqnarray}
where $\chi_{\sigma}^{c1,c2}=\chi_{xx}^{\sigma}\pm i\chi_{xy}^{\sigma}\mp i\chi_{yx}^{\sigma}+\chi_{yy}^{\sigma}$, again due to the fact that the electric field does not mix up the two spin species. Subtracting the absorption rates of the two protocols for each spin species, i.e., performing a spin-resolved circular dichroism measurement, yields 
\begin{eqnarray}
R_{\sigma}^{c1}({\bf k},\omega)-R_{\sigma}^{c2}({\bf k},\omega)
=2\pi\left(\frac{qE_{0}}{\hbar}\right)^{2}\,2\Omega^{d\sigma}_{xy}({\bf k},\omega).
\end{eqnarray}
In the last step, we subtract the results for the two spin species to obtain the dressed spin Berry curvature
\begin{eqnarray}
\sum_{\sigma}\frac{\sigma}{2}\left[R_{\sigma}^{c1}({\bf k},\omega)-R_{\sigma}^{c2}({\bf k},\omega)\right]
=2\pi\left(\frac{qE_{0}}{\hbar}\right)^{2}\,2\Omega^{ds}_{xy}({\bf k},\omega),
\end{eqnarray}
In particular, the sign of the Pfaffian of the ${\mathbb Z}_{2}$ invariant can be measured directly by probing the sign of the spin Berry curvature at HSPs according to Eq.~(\ref{sign_Pfaffian_sign_spin_Berry}), which may be achieved by simply measuring which spin species has more circular dichroism at HSPs in the pump-probe type of experiment using spin- and time-resolved ARPES, as demonstrated schematically in Fig.~\ref{fig:circ_dic_spin_Berry}.

\subsection{BHZ model in a continuum with impurity scattering}

\begin{figure}
\begin{center}
\includegraphics[clip=true,width=0.7\columnwidth]{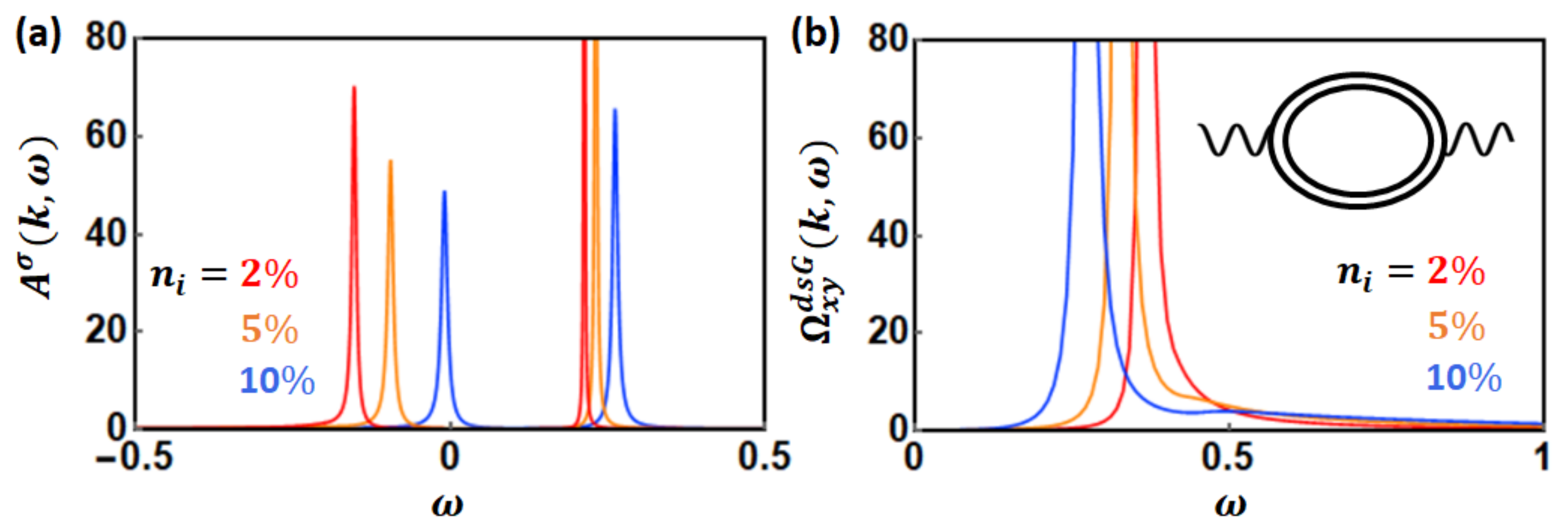}
\caption{(a) The spectral function of the linearized BHZ model at momentum $k=0.001$ and mass term $M=0.2$ at different impurity density $n_{i}$, which is the same for both spin species. The impurity potential is fixed at $V=1$. (b) The resulting spin Berry curvature spectral function $\Omega_{xy}^{dsG}({\bf k},\omega)$ at temperature $T=0.03$ calculated within the full Green's function approach indicated by the Feynman diagram in the inset, where the double lines are the full Green's functions and wavy lines the external electric field. The frequency integration of the three spectral functions yields spin Berry curvature $11.93$, $11.77$, and $10.58$ for impurity density $n_{i}=0.02$, $0.05$, and $0.1$, respectively, which are to be compared with the nonintercting zero temperature value $\Omega_{xy}^{s}({\bf k})=12.50$, indication the suppression of spin Berry curvature by impurities.   } 
\label{fig:spin_Berry_specfn_BHZ}
\end{center}
\end{figure}

Our linear response theory makes it possible to express spin Berry curvature in terms of Feynman diagrams, and hence calculate it by standard many-body techniques such as Dyson's equation, Bethe-Salpeter equation, frequency sum, etc., as have been demonstrated for the spinless version\cite{Chen22_dressed_Berry_metric}. As a concrete example, we consider the BHZ model in Eq.~(\ref{BHZ_Dirac_Hamiltonian}) linearized around the origin ${\bf k}=(0,0)$, described by the components $d_{1}=vk_{x}$, $d_{2}=vk_{y}$, and $d_{3}=M$, where $v=1$ is the Fermi velocity, and consider the effect of impurity scattering. We will consider the impurity potential $V\times I_{4\times 4}$ that causes only intraband scattering and does not mix up the two spin species 
\begin{eqnarray}
&&V_{\bf kk'}^{n\sigma}=\langle n{\bf k'}\sigma|V|n{\bf k}\sigma\rangle
=\frac{V}{2d(d-d_{3})}\left[(d_{3}-d)^{2}+(d_{1}^{2}+d_{2}^{2})e^{i\sigma(\varphi-\varphi ')}\right],
\nonumber \\
&&V_{\bf kk'}^{m\sigma}=\langle m{\bf k'}\sigma|V|m{\bf k}\sigma\rangle
=\frac{V}{2d(d+d_{3})}\left[(d_{3}+d)^{2}+(d_{1}^{2}+d_{2}^{2})e^{i\sigma(\varphi-\varphi ')}\right],
\end{eqnarray}
where $\varphi$ is the planar angle of the momentum ${\bf k}=(k,\varphi)$. 
The self-energy at an impurity density $n_{i}$ is calculated within the $T$-matrix approximation $\Sigma_{n/m}^{\sigma}({\bf k},\omega)=n_{i}T_{\bf kk}^{\sigma n/m}(\omega)$, which is then put into the Dyson's equation $G^{\sigma}=G^{\sigma(0)}+G^{\sigma(0)}\Sigma^{\sigma} G^{\sigma}$ to obtain the full Green's function\cite{Chen22_dressed_Berry_metric}. We then calculate the spin-selected charge polarization susceptibility within the full Green's function approximation described by the Feynman diagram in Fig.~\ref{fig:spin_Berry_specfn_BHZ} (b), yielding
\begin{eqnarray}
\chi_{xy}^{\sigma G}({\bf k},i\omega)
=\sum_{nm}{\cal A}_{x}^{\sigma nm}\left[{\cal A}_{y}^{\sigma nm}\right]^{\dag}
\frac{1}{\beta}\sum_{ip}G_{n}^{\sigma}({\bf k},ip)G_{m}^{\sigma}({\bf k},i\omega+ip).
\label{chi_fullG}
\end{eqnarray}
The single-particle spectral function of the full Green's function $A^{\sigma}({\bf k},\omega)=-{\rm Im}\,G^{\sigma}({\bf k},\omega)/\pi$ is then used to rewrite the Berry curvature of spin $\sigma$ by\cite{Mahan00}
\begin{eqnarray}
&&\Omega_{xy}^{d\sigma G}({\bf k},\omega)
=\Omega_{xy}^{\sigma}({\bf k})\int d\varepsilon A_{n}^{\sigma}({\bf k},\varepsilon)A_{m}^{\sigma}({\bf k},\varepsilon+\omega)\left[f({\varepsilon})-f({\varepsilon+\omega})\right],
\label{Berry_AA_formula}
\end{eqnarray} 
where $\Omega_{xy}^{\sigma}({\bf k})=i{\cal A}_{x}^{\sigma nm}\left[{\cal A}_{y}^{\sigma nm}\right]^{\dag}-i{\cal A}_{y}^{\sigma nm}\left[{\cal A}_{x}^{\sigma nm}\right]^{\dag}=\sigma\varepsilon^{abc}d_{a}\partial_{x}d_{b}\partial_{y}d_{c}/2$ is the noninteracting Berry curvature.

Figure \ref{fig:spin_Berry_specfn_BHZ} (a) shows the sum of the valence band and conduction band single-particle spectral functions $A^{\sigma}({\bf k},\omega)=A_{n}^{\sigma}({\bf k},\omega)+A_{m}^{\sigma}({\bf k},\omega)$ near the origin ${\bf k}=(0,0)$ for either spin species, where we fix the impurity potential $V=1$ and examine different impurity densities $n_{i}$. The width of quasiparticle peaks broadens at larger $n_{i}$, signifying reduced quasiparticle lifetime as expected. The peak positions imply that the band gap is reduced at large $n_{i}$. As a result, the maximum of the spin Berry curvature spectral function $\Omega_{xy}^{dsG}({\bf k},\omega)$ shown in Fig.~\ref{fig:spin_Berry_specfn_BHZ} (b) is shifted to smaller frequency at larger $n_{i}$ since it measures the excitation across the band gap. In addition, the broadening of the single-particle spectral function causes the broadening of the spin Berry curvature spectral function, owing to the fact that the optical absorption is no longer restricted at a single frequency. Numerically integrating the spectral function over frequency yields a spin Berry curvature that decreases with $n_{i}$, indicating the suppression of spin Berry curvature by impurity scattering. This result suggests that although topological order is generally considered to be protected by the band gap against various perturbations, in reality this is true only if the perturbation is weak enough. A strong perturbation can still significantly modify the profile of spin Berry curvature.


\section{Measurement in real space}

\subsection{Spin Chern marker \label{sec:spin_Chern_marker}}

Having done the momentum space formalism in the previous section, we proceed to discuss the spin Chern marker as a real space ${\mathbb Z}_{2}$ invariant for 2D class AII materials\cite{Prodan10_2,Prodan11}. We will follow the formalism of the (spinless) Chern marker originally proposed by Bianco and Resta\cite{Bianco11}, emphasize the connection to Wannier states, and apply a recently proposed linear response theory to connect the spin Chern marker to circular dichroism experiments. We start by denoting $\langle{\bf r}|\ell{\bf k}\sigma\rangle=u_{\ell{\bf k}\sigma}({\bf r})=e^{-i{\bf k\cdot r}}\psi_{\ell{\bf k}\sigma}({\bf r})$ as the periodic part of the Bloch state of spin $\sigma$ satisfying $u_{\ell{\bf k}\sigma}({\bf r})=u_{\ell{\bf k}\sigma}({\bf r+R})$, with ${\bf R}$ a Bravais lattice vector, and the band index $\ell$ refers to either the conduction or valence band. The Bloch state $|\ell{\bf k}\sigma\rangle$ defines a Wannier state $|{\bf R}\ell\sigma\rangle$ by (ignoring the normalization factor)
\begin{eqnarray}
|\ell{\bf k}\sigma\rangle=\sum_{{\bf R}}e^{-i {\bf k}\cdot({\hat{\bf r}}-{\bf R})}|{\bf R}\ell\sigma\rangle,\;\;\;
|{\bf R} \ell\sigma\rangle=\sum_{\bf k}e^{i {\bf k}\cdot({\hat{\bf r}}-{\bf R})}|\ell{\bf k}\rangle,
\label{Wannier_basis}
\end{eqnarray}
where ${\hat{\bf r}}$ is the position operator. Correspondingly, the Wannier function at position ${\bf r}=(x,y)$ that localizes around the home cell ${\bf R}$ is given by $\langle {\bf r}|{\bf R}\ell\sigma\rangle=W_{\ell\sigma}({\bf r}-{\bf R})$.

The quantity of our interest is the spin Chern number calculated from the momentum space integration ${\cal C}^{s}=\int\frac{d^{2}{\bf k}}{(2\pi)^{2}}\Omega_{xy}^{s}({\bf k})$ of the spin Berry curvature defined in Eq.~(\ref{2DclassAII_TrQdQ_PxQy}). The ${\mathbb Z}_{2}$ invariant is determined from the spin Chern number by $(-1)^{{\cal C}^{s}}$, since the spin Chern number is equivalent to the wrapping number ${\rm deg}[{\bf n}]$ as shown in Sec.~\ref{sec:spin_Berry_Jacobian}. Denoting the position operator of the whole lattice by ${\hat x}=\sum_{\bf r\sigma}|{\bf r\sigma}\rangle x_{\bf r}\langle{\bf r\sigma}|\equiv\sum_{\bf r}|{\bf r}\rangle x_{\bf r}\langle{\bf r}|$, and using the identity 
\cite{Bianco11,Marzari97,Marzari12,Gradhand12,Chen17,Chen19_AMS_review}
\begin{eqnarray}
&&i\langle m{\bf k}\sigma|\partial_{x}|n{\bf k}\sigma\rangle=\langle \psi_{m{\bf k}\sigma}|{\hat x}|\psi_{n{\bf k}\sigma}\rangle
\nonumber \\
&&=\sum_{\bf R}e^{i{\bf k\cdot R}}\langle{\bf 0}m\sigma|{\hat x}|{\bf R}n\sigma\rangle=\sum_{\bf R}e^{-i{\bf k\cdot R}}\langle{\bf R}m\sigma|{\hat x}|{\bf 0}n\sigma\rangle,
\label{udu_to_psixpsi}
\end{eqnarray}
valid for different states $m\neq n$ but with the same spin $\sigma$, one can express the spin Chern number\cite{Sheng06,Prodan07,Fukui07} given by integrating Eq.~(\ref{2DclassAII_TrQdQ_PxQy}) in terms of Wannier states\cite{Bianco11}
\begin{eqnarray}
&&{\cal C}^{s}=\sum_{nm\sigma}\int\frac{d^{2}{\bf k}}{(2\pi)^{2}}\frac{i}{2}\langle\psi_{n{\bf k}\sigma}|{\hat x}{\hat\sigma}^{z}|\psi_{m{\bf k}\sigma}\rangle\langle \psi_{m{\bf k}\sigma}|{\hat y}|\psi_{n{\bf k}\sigma}\rangle-( x\leftrightarrow y)
\nonumber \\
&&=\sum_{nm\sigma}\int\frac{d^{2}{\bf k}}{(2\pi)^{2}}\int\frac{d^{2}{\bf k'}}{(2\pi)^{2}}\frac{i}{2}\langle\psi_{n{\bf k}\sigma}|{\hat x}{\hat\sigma}^{z}|\psi_{m{\bf k'}\sigma}\rangle\langle \psi_{m{\bf k'}\sigma}|{\hat y}|\psi_{n{\bf k}\sigma}\rangle-( x\leftrightarrow y)
\nonumber \\
&&=\sum_{nm\sigma}\sum_{\bf R}\frac{i}{2}\langle{\bf 0}n\sigma|{\hat x}{\hat\sigma}^{z}|{\bf R}m\sigma\rangle\langle{\bf R}m\sigma|{\hat y}|{\bf 0}n\sigma\rangle-( x\leftrightarrow y)
\nonumber \\
&&=\sum_{nm\sigma}\sum_{\bf R}\frac{i}{2}\int d{\bf r}\int d{\bf r'}\,\sigma x_{\bf r}W_{n\sigma}^{\ast}({\bf r})W_{m\sigma}({\bf r-R})y_{\bf r'}W_{m\sigma}^{\ast}({\bf r'-R})W_{n\sigma}({\bf r'})
\nonumber \\
&&-( x\leftrightarrow y).
\label{Cval_Tr_cell}
\end{eqnarray}
Introducing the projection operator to the valence band states, and to the spin-selected conduction band states 
\begin{eqnarray}
{\hat P}=\sum_{n\sigma}\int\frac{d^{2}{\bf k}}{(2\pi)^{2}}|\psi_{n{\bf k}\sigma}\rangle\langle\psi_{n{\bf k}\sigma}|,\;\;\;{\hat Q}=\sum_{m\sigma}\int\frac{d^{2}{\bf k'}}{(2\pi)^{2}}|\psi_{m{\bf k'}\sigma}\rangle\langle\psi_{m{\bf k'}\sigma}|,
\label{projector_PQ_k}
\end{eqnarray}
the spin Chern number may be written as
\begin{eqnarray}
&&{\cal C}^{s}=\frac{i}{2}{\rm Tr}_{\rm cell}\left[{\hat P}{\hat x}{\hat\sigma}^{z}{\hat Q}{\hat  y}-{\hat P}{\hat y}{\hat\sigma}^{z}{\hat Q}{\hat  x}\right].
\end{eqnarray}
We will call $i\left[{\hat P}{\hat x}{\hat\sigma}^{z}{\hat Q}{\hat  y}-{\hat P}{\hat y}{\hat\sigma}^{z}{\hat Q}{\hat  x}\right]/2$ the spin Chern operator.

\begin{figure}[ht]
\begin{center}
\includegraphics[clip=true,width=0.7\columnwidth]{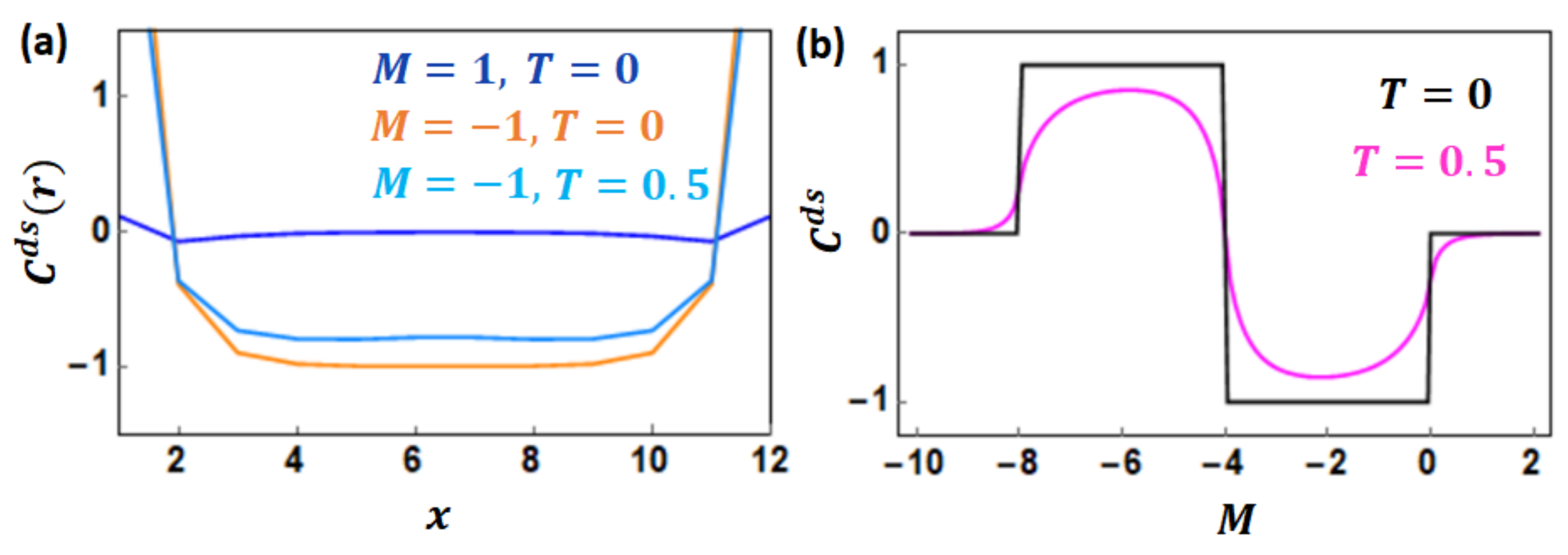}
\caption{(a) The dressed spin Chern marker ${\cal C}^{ds}({\bf r})={\cal C}^{ds}(x,y)$ along the center line $y=6$ of a $12\times 12$ lattice BHZ model evaluated at zero and finite temperatures, and at different values of the mass term $M$ that controls the topological order. The zero temperature result is equivalently denoted by ${\cal C}^{ds}({\bf r})|_{T=0}={\cal C}^{s}({\bf r})$. (b) The dressed spin Chern number of the BHZ model in the thermodynamic limit as a function of the mass term $M$ at zero and a finite temperature. } 
\label{fig:BHZ_Cds_data}
\end{center}
\end{figure}

For a 2D lattice model that has been diagonalized $H|E_{n}\rangle=E_{n}|E_{n}\rangle$, we generalize the projectors in Eq.~(\ref{projector_PQ_k}) to\cite{Bianco11} 
\begin{eqnarray}
{\hat P}=\sum_{n}|E_{n}\rangle\langle E_{n}|,\;\;\;{\hat Q}=\sum_{m}|E_{m}\rangle\langle E_{m}|.
\label{projector_PQ_En}
\end{eqnarray}
from which the local spin Chern marker at ${\bf r}$ can be introduced\cite{Bianco11}
\begin{eqnarray}
&&{\cal C}^{s}({\bf r})=\frac{i}{2}\sum_{\sigma}\langle{\bf r\sigma}|\left[{\hat P}{\hat x}{\hat\sigma}^{z}{\hat Q}{\hat  y}-{\hat P}{\hat y}{\hat\sigma}^{z}{\hat Q}{\hat  x}\right]|{\bf r\sigma}\rangle\equiv \frac{i}{2}\langle{\bf r}|\left[{\hat P}{\hat x}{\hat\sigma}^{z}{\hat Q}{\hat  y}-{\hat P}{\hat y}{\hat\sigma}^{z}{\hat Q}{\hat  x}\right]|{\bf r}\rangle
\nonumber \\
&&=\sum_{nm\sigma}\sum_{\bf R}\frac{i}{2}\langle{\bf 0}n\sigma|{\hat x}_{\bf r}{\hat \sigma}^{z}|{\bf R}m\sigma\rangle\langle{\bf R}m\sigma|{\hat y}|{\bf 0}n\sigma\rangle-( x\leftrightarrow y)
\nonumber \\
&&=\sum_{nm}\frac{i}{2}\langle E_{n}|{\hat x}_{\bf r}{\hat\sigma}^{z}|E_{m}\rangle\langle E_{m}|{\hat y}|E_{n}\rangle
-( x\leftrightarrow y).
\label{Cr_def_1}
\end{eqnarray}
Numerical result for ${\cal C}^{s}({\bf r})$ along the center line of a $12\times 12$ lattice BHZ model in the periodic boundary condition (PBC) is shown in Fig.~\ref{fig:BHZ_Cds_data} (a) for different values of the mass term $M$ that controls the topological order. In a large region near the center of the lattice, ${\cal C}^{s}({\bf r})$ well reproduces the spin Chern number expected for the BHZ model (${\cal C}^{s}({\bf r})\approx -1$ for $M=-1$ and ${\cal C}^{s}({\bf r})\approx 0$ for $M=1$), pointing to the validity of our formalism. Near the boundary of the lattice, ${\cal C}^{s}({\bf r})$ starts to deviate from the expected value, a problem known for this kind of construction\cite{Bianco11} that can be fixed by a more sophisticated treatment on the position operators\cite{Prodan10,Prodan11}. Nevertheless, ${\cal C}^{s}({\bf r})$ near the center of the lattice serves as a faithful index to map out the topological invariant of 2D class AII systems, suggesting that it can further be used to investigate the effect of real space inhomogeneity such as disorder and grain boundary. As we aim to focus on the spectroscopic aspects related to the measurement of the spin Chern marker ${\cal C}^{s}({\bf r})$, which will be formulated in the following sections, the effect of inhomogeneity will be left for future investigations.

\subsection{Linear response theory and measurement of spin Chern marker}

Similar to that discussed in Sec.~\ref{sec:linear_response_kspace}, below we show that the spin Chern marker can also be measured by spin-resolved circular dichroism in real space. Consider the zero temperature case first. Suppose one applies the two circularly polarized electric field described by Eq.~(\ref{hc1c2_kspace}), but now written in real space for a lattice model
\begin{eqnarray}
H_{c1,c2}=({\hat x}+i{\hat y})qE_{0}e^{-i\omega t},
\end{eqnarray}
which again does not mix up the two spin species. Labeling the spin eigenvalues $\sigma$ in the eigenstates $|E_{\ell\sigma}\rangle$ explicitly, the Fermi Golden rule yields the total transition rate from all the occupied states $|E_{n\sigma}\rangle$ of spin $\sigma$ to all the empty states $|E_{m\sigma}\rangle$ with the same spin by
\begin{eqnarray}
\Delta n_{\sigma}^{c1,c2}(\omega)=\frac{2\pi}{\hbar}\sum_{nm}|\langle E_{m\sigma}|H_{c1,c2}|E_{n\sigma}\rangle|^{2}\delta(\hbar\omega+E_{n\sigma}-E_{m\sigma}),
\end{eqnarray}
The spin Chern number ${\cal C}^{s}$ is then given by the difference between the circular dichroism of the two spin species integrated over frequency
\begin{eqnarray}
\int d\omega\sum_{\sigma}\frac{\sigma}{2}\left(\Delta n_{\sigma}^{c1}(\omega)-\Delta n_{\sigma}^{c1}(\omega)\right)=4\pi A\left(\frac{qE_{0}}{\hbar}\right)^{2}{\cal C}^{s},
\label{Deltan_Chern_number}
\end{eqnarray}
as in the spinless version\cite{Molignini22_Chern_marker}, where $A$ is the unit cell area. 


\begin{figure}[ht]
\begin{center}
\includegraphics[clip=true,width=0.7\columnwidth]{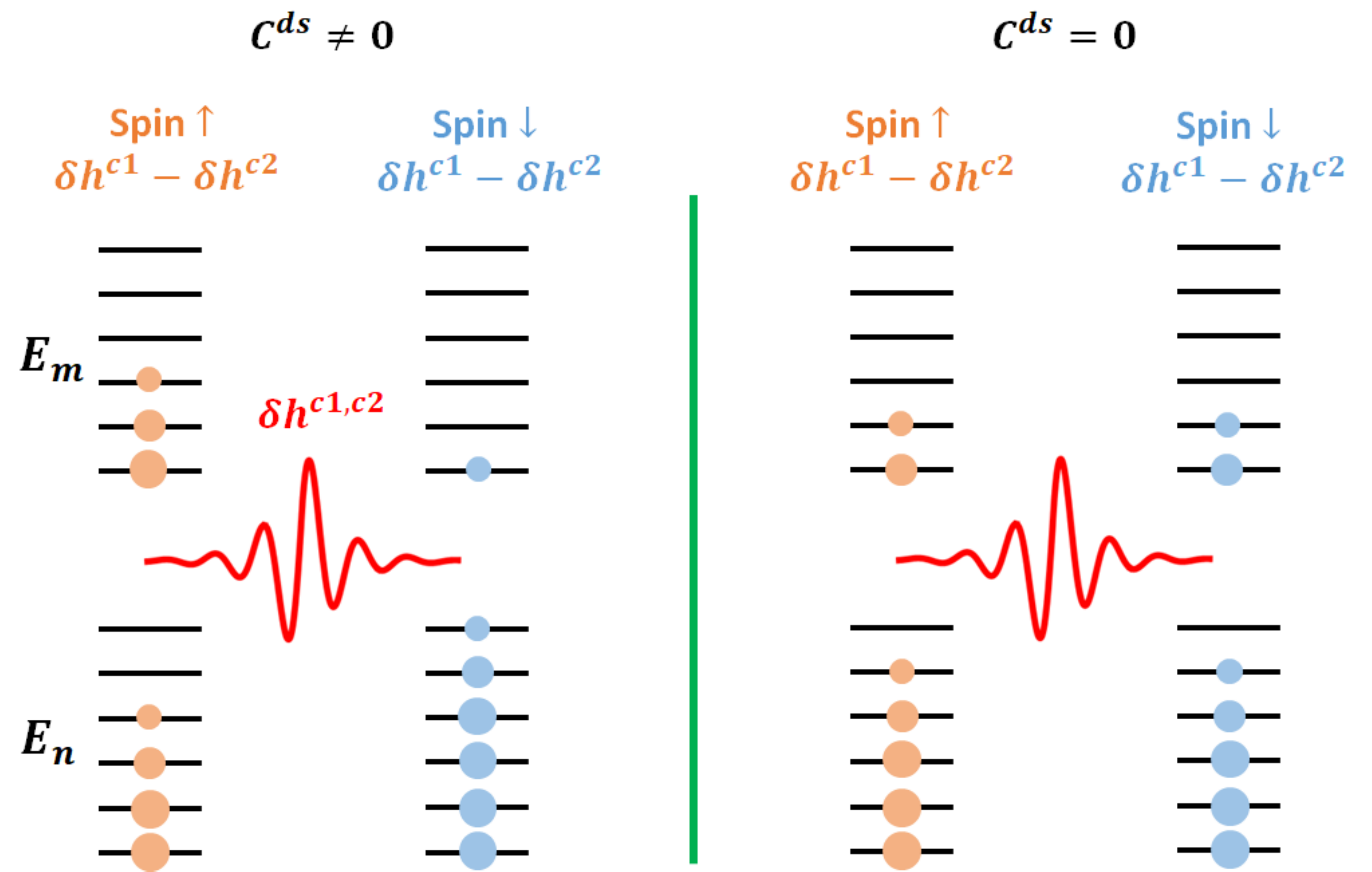}
\caption{The proposed real space experiment to measure whether the dressed spin Chern number ${\cal C}^{ds}$ is nonzero. In the spin-resolved circular dichroism experiment, if one spin species excited from filled states $E_{n}$ to the empty states $E_{m}$ is more than the other spin species then ${\cal C}^{ds}\neq 0$ (left panel), but if the two spin species are excited by the same amount then ${\cal C}^{ds}=0$ (right panel). The filling of spin species is indicated by colored points, which may be measured by time- and spin-resolved photoemission. The local spin Chern marker ${\cal C}^{ds}({\bf r})$ at position ${\bf r}$ can be measured if the photoemission has a spatial resolution, whose spectral function is denoted by ${\cal C}^{ds}({\bf r},\omega)$. } 
\label{fig:circ_dic_spin_Chern_marker}
\end{center}
\end{figure}

The measurement protocol can be generalized to finite temperature by means of a linear response theory. Suppose one applies a uniform electric field $E^{ y}$ oscillating with frequency $\omega$ along direction ${\hat{\bf y}}$, then the expectation value of ${\hat x}_{\bf r}(t)$ of spin $\sigma$ induced by the field defines a charge polarization susceptibility $\chi_{ x y}^{\sigma}$ at ${\bf r}$ for spin $\sigma$
\begin{eqnarray}
\langle {\hat x}_{\bf r}^{\sigma}(t)\rangle_{0}=\chi_{ x y}^{\sigma}({\bf r},t)\left[-qE^{ y}(t)\right]
=-\chi_{ x y}^{\sigma}({\bf r},t)qE^{ y}e^{-i\omega t},
\end{eqnarray}
The corresponding nonlocal charge polarization response is given by
\begin{eqnarray}
\chi_{ x y}^{\sigma}({\bf r,r',}t-t')=-i\theta(t-t')\langle {\hat x}_{\bf r}^{\sigma}(t){\hat y}_{\bf r'}^{\sigma\dag}(t')-{\hat y}_{\bf r'}^{\sigma\dag}(t'){\hat x}_{\bf r}^{\sigma}(t)\rangle,
\end{eqnarray}
The usual procedure in linear response theory gives the retarded nonlocal response 
\begin{eqnarray}
\chi_{xy}^{\sigma}({\bf r,r',}\omega)=\sum_{\ell\ell '}\langle E_{\ell\sigma}|{\hat x}_{\bf r}|E_{\ell '\sigma}\rangle\langle E_{\ell '\sigma}|{\hat y}_{\bf r'}|E_{\ell\sigma}\rangle\frac{f(E_{\ell\sigma})-f(E_{\ell '\sigma})}{\omega+E_{\ell\sigma}-E_{\ell '\sigma}+i\eta},
\label{chi_rrp}
\end{eqnarray}
where $f(E_{\ell\sigma})$ is the Fermi function and $\eta$ is a small artificial broadening. Note that $\ell$ and $\ell '$ both enumerate the filled and empty states. The local susceptibility $\chi_{ x y}^{\sigma}({\bf r,}\omega)=\sum_{\bf r'}\chi_{ x y}^{\sigma}({\bf r,r',}\omega)$ at site ${\bf r}$ is given by summing over the nonlocal susceptibility between ${\bf r}$ and all the sites on the lattice ${\bf r'}$.

The finite temperature Chern number spectral function for spin $\sigma$ at positive frequency $\omega>0$ is defined by
\begin{eqnarray}
&&{\cal C}^{d\sigma}(\omega)=-\frac{1}{\pi}{\rm Re}\sum_{\bf r}\left[\chi_{ x  y}^{\sigma}({\bf r},\omega)-\chi_{ y x}^{\sigma}({\bf r},\omega)\right]
\nonumber \\
&&={\rm Tr}\sum_{\ell<\ell '}\left[i\langle E_{\ell\sigma}|{\hat x}|E_{\ell '\sigma}\rangle\langle E_{\ell '\sigma}|{\hat y}|E_{\ell\sigma}\rangle-( x\leftrightarrow y)\right]
\nonumber \\
&&\times\left[f(E_{\ell\sigma})-f(E_{\ell '\sigma})\right]\delta(\omega+E_{\ell\sigma}-E_{\ell '\sigma}),
\label{total_Chern_spec_fn}
\end{eqnarray}
where $\ell<\ell '$ means $E_{\ell}<E_{\ell '}$. The difference between spin up and down is referred to as the dressed spin Chern number spectral function ${\cal C}^{ds}(\omega)=\left[{\cal C}^{d\uparrow}(\omega)-{\cal C}^{d\downarrow}(\omega)\right]/2$, whose frequency integration gives the dressed spin Chern number at finite temperature $\int_{0}^{\infty}d\omega\,{\cal C}^{ds}(\omega)={\cal C}^{ds}$. We anticipate that ${\cal C}^{ds}(\omega)$ may be measurable by performing pump-probe type of measurement in real space using circularly polarized light, as elaborated in Fig.~\ref{fig:circ_dic_spin_Chern_marker}. The two spin species are expected to show a difference in circular dichroism in topologically nontrivial phases, and no difference in the topologically trivial phases, which should be detectable by time- and spin-resolved photoemission.

One can further write the above spectral function as the sum of individual site ${\cal C}^{d\sigma}(\omega)=\sum_{\bf r}{\cal C}^{d\sigma}({\bf r},\omega)$, which introduces the local Chern marker spectral function for spin $\sigma$
\begin{eqnarray}
{\cal C}^{d\sigma}({\bf r},\omega)&=&\sum_{\ell<\ell '}{\rm Re}\left[i\langle E_{\ell\sigma}|{\hat x}_{\bf r}|E_{\ell '\sigma}\rangle\langle E_{\ell '\sigma}|{\hat y}|E_{\ell\sigma}\rangle-( x\leftrightarrow y)\right]
\nonumber \\
&&\times\left[f(E_{\ell\sigma})-f(E_{\ell '\sigma})\right]\delta(\omega+E_{\ell\sigma}-E_{\ell '\sigma}).
\label{generalized_Chern_marker}
\end{eqnarray}
Similar to that introduced above, the difference between spin up and down gives the spin Chern marker spectral function ${\cal C}^{ds}({\bf r},\omega)=\left[{\cal C}^{d\uparrow}({\bf r},\omega)-{\cal C}^{d\downarrow}({\bf r},\omega)\right]/2$, and a frequency integration gives the spin Chern marker at finite temperature $\int_{0}^{\infty}d\omega\,{\cal C}^{ds}({\bf r},\omega)={\cal C}^{ds}({\bf r})$. The numerical calculation of the Fermi function factor and the $\delta$-function in Eq.~(\ref{generalized_Chern_marker}) can be conveniently interpreted by the method discussed in Ref.~\cite{Molignini22_Chern_marker}. Note that in the zero temperature limit $T\rightarrow 0$, the Fermi function factor $\left[f(E_{\ell})-f(E_{\ell '})\right]$ just picks up filled $\ell =n\in v$ and empty $\ell '=m\in c$ states, and hence after an integration over frequency ${\cal C}^{ds}({\bf r})|_{T=0}={\cal C}^{s}({\bf r})$, one recovers the zero temperature spin Chern marker in Eq.~(\ref{Cr_def_1}). Moreover, the $\left[f(E_{\ell})-f(E_{\ell '})\right]$ factor at finite temperature is always less than unity, making ${\cal C}^{ds}({\bf r})$ smaller than its zero temperature value, as can be seen in Fig.~\ref{fig:BHZ_Cds_data} (a).

An analytical expression of the spectral function can be obtained for linear Dirac Hamiltonians. Focusing on the $M=0$ critical point of the BHZ model in Eq.~(\ref{BHZ_Dirac_Hamiltonian}) caused by gap closing at ${\bf k}_{0}=(0,0)$, one can linearize the Hamiltonian around this poiint to obtain $d_{1}=vk_{x}$, $d_{2}=vk_{y}$, $d_{3}=M$. The spin Chern number/marker spectral function in this homogeneous, continuous model is given by
\begin{eqnarray}
{\cal C}^{ds}(\omega)=\int\frac{d^{2}{\bf k}}{(2\pi)^{2}}\sum_{\sigma}\frac{\sigma}{2}\,\Omega_{xy}^{\sigma}\left[f(E_{n\sigma})-f(E_{m\sigma})\right]
\delta(\omega+E_{n\sigma}-E_{m\sigma}).
\label{Csw_BHZ_specfun}
\end{eqnarray}
The result is completely the same as the Chern insulator that has been discussed previously: At zero temperature ${\cal C}^{ds}(\omega)\rightarrow {\cal C}^{s}(\omega)$, the spectral function is finite only at frequency larger than the bulk gap $\omega>2|M|/\hbar$, and scales like ${\cal C}^{s}(\omega)\sim M/\omega^{2}$ such that the integration over frequency gives the topological invariant. 


\begin{figure}[ht]
\begin{center}
\includegraphics[clip=true,width=0.7\columnwidth]{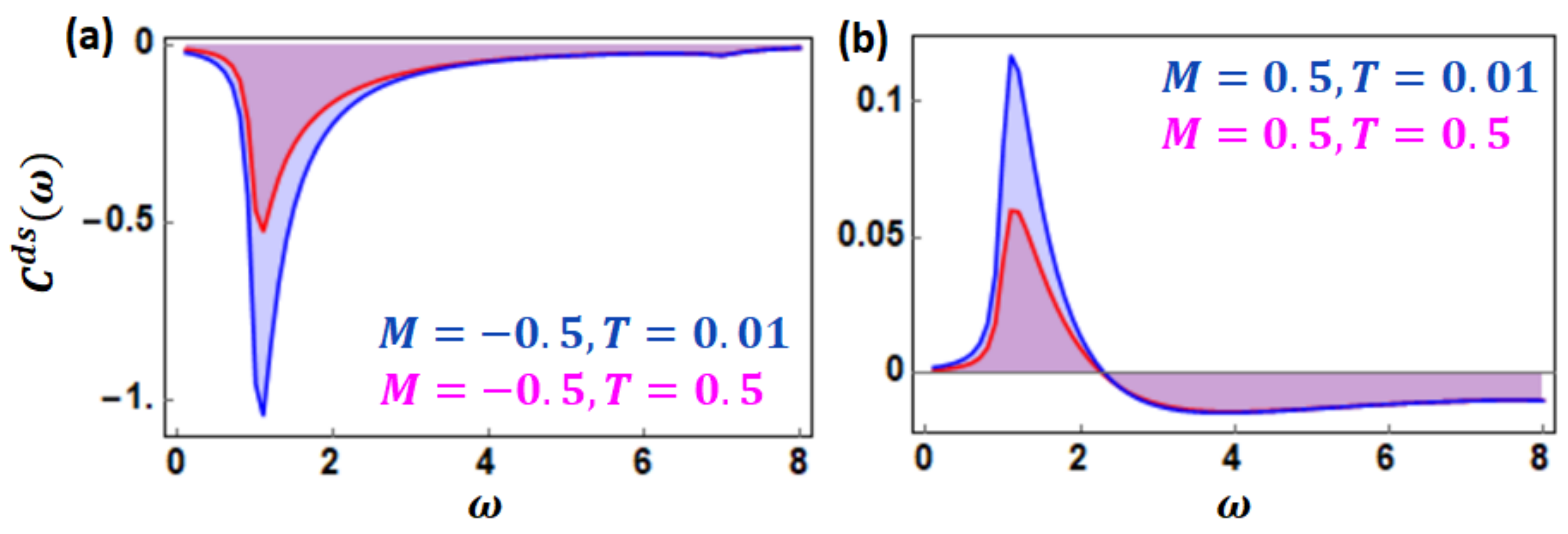}
\caption{The spin Chern number/marker spectral function of the BHZ model in the (a) topologically nontrivial and (b) trivial phases, evaluated at both zero $T=0$ and a finite temperature $T=0.5$. } 
\label{fig:BHZ_specfn}
\end{center}
\end{figure}

The Chern number/marker spectral function ${\cal C}^{ds}(\omega)$ for the homogeneous lattice BHZ model in Eq.~(\ref{BHZ_Dirac_Hamiltonian}) is shown in Fig.~\ref{fig:BHZ_specfn} for $M=\pm 0.5$ at a low $T=0.01$ and a high temperature $T=0.5$, which displays the following features. Firstly, the spectral funtion inside the band gap $\omega\apprle 2|M|$ is practically zero, agree with that expected from an exciton absorption rate. A large spectral weight is centering at the frequency close to the band gap $\omega\apprge 2|M|$, indicating that the states near the band gap are the most detrimental to the topological properties. Comparing the $M=0.5$ and $M=-0.5$ results at low temperature $T=0.01$, it is also clear that the sign change of the this low frequency part near the band gap causes the frequency integration of the spectral function to change from zero to finite, in agreement with the usual picture that band inversion causes a TPT. On the other hand, the effect of finite temperature is to reduce the spectral weight at low energy, as expected since the Fermi function factor $f(E_{n\sigma})-f(E_{m\sigma})$ in Eq.~(\ref{Csw_BHZ_specfun}) mainly influences the low energy states in practice. As a consequence, in the topologically nontrivial phase $M=-0.5$, the frequency integrated spin Chern number will be reduced $|{\cal C}^{ds}|<1$, whereas in the topologically trivial phase $M=0.5$ it will not be entirely zero $|{\cal C}^{ds}|>0$. This explains the ${\cal C}^{ds}$ versus $M$ profile at finite temperature presented Fig.~\ref{fig:BHZ_Cds_data} (b), as well as the smearing of the discrete jump of ${\cal C}^{ds}$ at critical points due to finite temperature. We anticipate that these generic features should also manifest in realistic materials, which can be easily verifiable by performing the proposed experiment on 2D materials that exhibit QSHE\cite{Yang12,Drozdov14,Tang17,Jia17,Peng17,Reis17,Fei17,Wu18,Deng18,Xu18,Zhu19,Shi19}.



\subsection{Topological quantum criticality}

From our spin Chern marker formalism, one can further construct two nonlocal objects that characterizes the quantum criticality near TPTs. The first is what we call the spin Chern correlator ${\tilde{\cal C}}^{s}({\bf r,r'})$ in the zero temperature limit, which is introduced by splitting the second position operator in Eq.~(\ref{Cr_def_1}) into its component at each lattice site ${\bf r}'$
\begin{eqnarray}
&&{\tilde{\cal C}}^{s}({\bf r,r'})={\rm Re}\langle{\bf r}|\left[\frac{i}{2}{\hat P}{\hat x}{\hat \sigma}^{z}{\hat Q}{\hat y}_{\bf r'}-\frac{i}{2}{\hat P}{\hat y}{\hat \sigma}^{z}{\hat Q}{\hat x}_{\bf r'}\right]|{\bf r}\rangle
\nonumber \\
&&=\sum_{nm\sigma}\sum_{\bf R}\frac{i}{2}\langle{\bf 0}n\sigma|{\hat x}_{\bf r}{\hat \sigma}^{z}|{\bf R}m\sigma\rangle\langle{\bf R}m\sigma|{\hat y}_{\bf r'}|{\bf 0}n\sigma\rangle-( x\leftrightarrow y)
\nonumber \\
&&=\sum_{nm}{\rm Re}\left\{\frac{i}{2}\langle E_{n}|{\hat x}_{\bf r}{\hat\sigma}^{z}|E_{m}\rangle\langle E_{m}|{\hat y}_{\bf r'}|E_{n}\rangle
-( x\leftrightarrow y)\right\},
\label{zeroT_Chern_correlator}
\end{eqnarray}
which sums to the spin Chern marker $\sum_{\bf r'}{\tilde{\cal C}}^{s}({\bf r,r'})={\cal C}^{s}({\bf r})$. This correlator decays with ${\bf r-r'}$ with a decay length $\xi$, and as the system approaches the critical point $M\rightarrow M_{c}$, the decay length $\xi$ gradually diverges but the overall magnitude of ${\tilde{\cal C}}^{s}({\bf r,r'})$ decreases such that the topological invariant $\sum_{\bf r'}{\tilde{\cal C}}^{s}({\bf r,r'})={\cal C}^{s}({\bf r})$ can remain constant, as has been demonstrated for the spinless version\cite{Molignini22_Chern_marker}. In addition, the finite temperature version of this correlator can be constructed from the linear response theory, which is nothing but the nonlocal charge polarization susceptibility. We call the spectral function of this correlator the spin Chern correlator spectral function
\begin{eqnarray}
&&{\tilde{\cal C}}^{ds}({\bf r,r'},\omega)=-\frac{1}{\pi}{\rm Re}\sum_{\sigma}\frac{\sigma}{2}\left[\chi_{xy}^{\sigma}({\bf r,r'},\omega)-\chi_{yx}^{\sigma}({\bf r,r'},\omega)\right]
\nonumber \\
&&=\sum_{\ell<\ell '}{\rm Re}\left[\frac{i}{2}\langle E_{\ell}|{\hat x}_{\bf r}{\hat\sigma}^{z}|E_{\ell '}\rangle\langle E_{\ell '}|{\hat y}_{\bf r'}|E_{\ell}\rangle-( x\leftrightarrow y)\right]
\nonumber \\
&&\times\left[f(E_{\ell})-f(E_{\ell '})\right]\delta(\omega+E_{\ell}-E_{\ell '}),
\label{spin_Chern_corr_spec_fn}
\end{eqnarray} 
and the correlator itself is obtained from the frequency integration of this object $\int_{0}^{\infty}d\omega\,{\tilde{\cal C}}^{ds}({\bf r,r'},\omega)={\tilde{\cal C}}^{ds}({\bf r,r'})$. Experimentally, ${\tilde{\cal C}}^{ds}({\bf r,r'},\omega)$ can be detected by applying the circularly polarized electric field of frequency $\omega$ at ${\bf r}$ and measuring the charge polarization of spin $\sigma$ at ${\bf r'}$, and then subtracting the result for the two spin species.

\begin{figure}[ht]
\begin{center}
\includegraphics[clip=true,width=0.99\columnwidth]{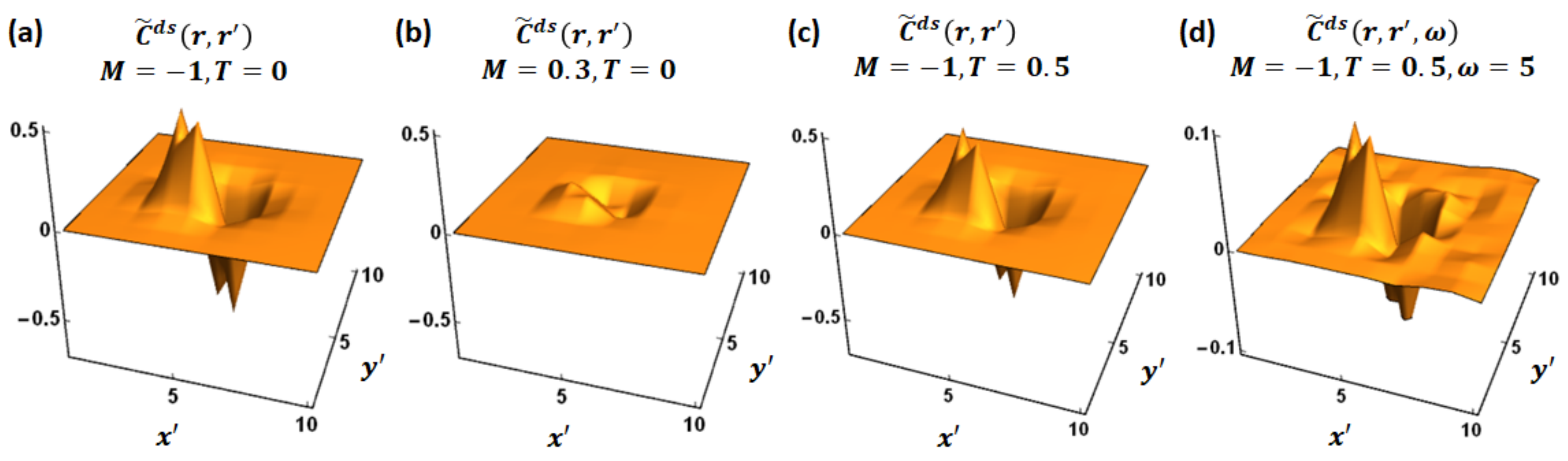}
\caption{The spin Chern correlator ${\tilde{\cal C}}^{ds}({\bf r,r'})$ for the lattice BHZ model of size $10\times 10$, with ${\bf r}$ fixed at the center of the lattice and plotted as a function of ${\bf r'}=(x',y')$. The spatial profile is evaluated at $T=0$ for (a) the topologically nontrivial phase $M=-1$ and (b) trivial phase $M=0.3$, and (c) for the nontrivial phase at $T=0.5$ whose spectral function at a specific frequency $\omega=5$ is shown in (d). } 
\label{fig:spin_Chern_correlator_data}
\end{center}
\end{figure}

The spatial profile of spin Chern marker ${\tilde{\cal C}}^{ds}({\bf r,r'})$ for the lattice BHZ model at zero temperature (equivalently ${\tilde{\cal C}}^{s}({\bf r,r'})$) is shown in Fig.~\ref{fig:spin_Chern_correlator_data} (a) and (b) for an example in the topologically nontrivial $M=-1$ and trivial $M=0.1$ phases, respectively, where we fix ${\bf r}$ to be at the center of the lattice and plot it as a function of ${\bf r}'$. Both cases show a spatial profile that contains positive and negative peaks, but the peaks have different weight in the nontrivial phase such that they sum to a finite Chern marker ${\tilde{\cal C}}^{ds}({\bf r})\approx -1$, whereas they have the same weight in the trivial phase such that ${\tilde{\cal C}}^{ds}({\bf r})\approx 0$. In addition, the magnitude of ${\tilde{\cal C}}^{ds}({\bf r,r'})$ is much larger in the nontrivial phase, indicating the the magnitude can be used to judge whether the system is moving towards a trivial or nontrivial phase upon tuning some system parameter. As depicted in Fig.~\ref{fig:spin_Chern_correlator_data} (c), the effect of finite temperature is to reduce the overall magnitude of these positive and negative peaks, a feature that is expected from the $f(E_{\ell})-f(E_{\ell '})$ factor in Eq.~(\ref{spin_Chern_corr_spec_fn}). On the other hand, the spin Chern correlator spectral function ${\tilde{\cal C}}^{ds}({\bf r,r'},\omega)$ displays a more complicated spatial pattern that depends on the frequency $\omega$ of the circularly polarized electric field, as shown in Fig.~\ref{fig:spin_Chern_correlator_data} (d) for an example, suggesting that a frequency integration of the proposed nonlocal circular dichroism experiment is necessary to obtain the peak-like profile of the correlator ${\tilde{\cal C}}^{ds}({\bf r,r'})$. Finally, we remark that these features are exactly the same as the spinless counterpart of the correlator in the Chern insulator that has been investigated in detail recently\cite{Molignini22_Chern_marker}, since the BHZ model is essentially two copies of the Chern insulator.


\begin{figure}[ht]
\begin{center}
\includegraphics[clip=true,width=0.7\columnwidth]{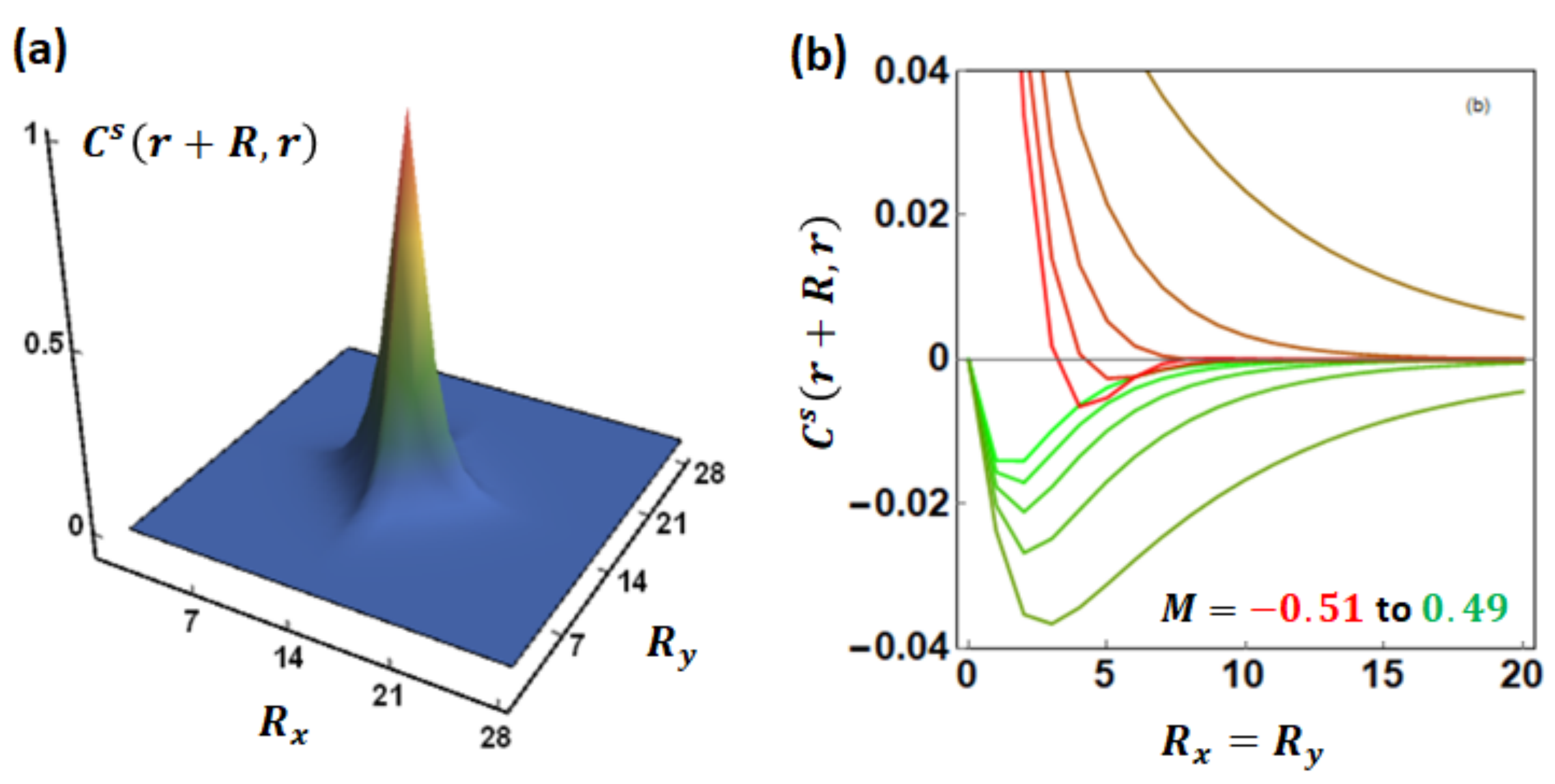}
\caption{(a) The nonlocal spin Chern marker ${\cal C}^{s}({\bf r+R,r})$ in a $28\times 28$ lattice BHZ model, where we choose ${\bf r}$ to be at the center of the lattice and plot it as a function of ${\bf R}=(R_{x},R_{y})$. (b) The nonlocal marker along the diagonal $R_{x}=R_{y}$ as the mass term $M$ is tuned across the critical point $M_{c}=0$.  } 
\label{fig:nonlocal_spin_Chern_data}
\end{center}
\end{figure}

The second object we introduced is the nonlocal spin Chern marker, which at zero temperature is a kind of spin-selected Wannier state correlation function\cite{Chen17,Chen19_AMS_review,Chen19_universality}, as we shall see below. Consider the Fourier transform of the spin Berry curvature expressed in terms of the spinful Wannier states\cite{Marzari12,Gradhand12,Wang06} 
\begin{eqnarray}
&&\tilde{F}^{s}({\bf R})=\int\frac{d^{2}{\bf k}}{(2\pi)^{2}}\Omega_{ x y}^{s}({\bf k})e^{i{\bf k\cdot R}}=-i\sum_{n\sigma}\frac{\sigma}{2}\langle{\bf R}n\sigma|(R_{x}\hat{y}-R_{y}\hat{x})|{\bf 0}n\sigma\rangle
\nonumber \\
&&=-i\sum_{n\sigma}\int d^{2}{\bf r}\,\frac{\sigma}{2}(R_{x}y-R_{y}x)W_{n\sigma}({\bf r-R})^{\ast}W_{n\sigma}({\bf r})
\end{eqnarray}
provide ${\bf R\neq 0}$. Moreover, we can further use the formalism in Sec.~\ref{sec:spin_Chern_marker} to write 
\begin{eqnarray}
&&\tilde{F}^{s}({\bf R})=\sum_{n}\int\frac{d^{2}{\bf k}}{(2\pi)^{2}}\frac{i}{2}\langle\partial_{ x} n{\bf k}|{\hat\sigma}^{z}|\partial_{ y}n{\bf k}\rangle e^{i{\bf k\cdot R}}-( x\leftrightarrow y)
\nonumber \\
&&=\sum_{n}\sum_{m}\int\frac{d^{2}{\bf k}}{(2\pi)^{2}}\int\frac{d^{2}{\bf k'}}{(2\pi)^{2}}\frac{i}{2}\langle\psi_{n{\bf k}}|{\hat x}{\hat\sigma}^{z}|\psi_{m{\bf k'}}\rangle\langle \psi_{m{\bf k'}}|{\hat y}|\psi_{n{\bf k}}\rangle e^{i{\bf k\cdot R}}-( x\leftrightarrow y).
\nonumber \\
\end{eqnarray}
Note that $|\psi_{n{\bf k}}\rangle e^{i{\bf k\cdot R}}$ projected to $\langle{\bf r}|$ is equal to $|\psi_{n{\bf k}}\rangle$ projected to $\langle{\bf r+R}|$. In terms of the projector algebra in Sec.~\ref{sec:spin_Chern_marker}, if we consider the $({\bf r+R,r})$-th off-diagonal element of the spin Chern operator and call it the nonlocal spin Chern marker ${\cal C}^{s}({\bf r+R,r})$
\begin{eqnarray}
&&{\cal C}^{s}({\bf r+R,r})\equiv \frac{i}{2}\langle{\bf r+R}|\left[{\hat P}{\hat x}{\hat \sigma}^{z}{\hat Q}{\hat  y}-{\hat P}{\hat y}{\hat \sigma}^{z}{\hat Q}{\hat  x}\right]|{\bf r}\rangle,
\end{eqnarray}
then the spin-selected Wannier state correlation function $\tilde{F}^{s}({\bf R})$ is the clean, theomodynamic limit of ${\cal C}^{s}({\bf r+R,r})$
\begin{eqnarray}
\lim_{N\rightarrow\infty}{\rm Re}\left[{\cal C}^{s}({\bf r+R,r})\right]=\tilde{F}^{s}({\bf R}).
\end{eqnarray}
Particularly for the BHZ model, the spin Berry curvature near HSPs as the system approaches TPTs, which dictates that ${\cal C}^{s}({\bf r+R,r})$ must decay with ${\bf R}$, as shown in Fig.~\ref{fig:nonlocal_spin_Chern_data} (a) for an example. The decay length $\xi$ is inversely proportional to the band gap, and hence diverges near TPT, as can be seen from the evolution of ${\cal C}^{s}({\bf r+R,r})$ across the $M_{c}=0$ critical point shown in Fig.~\ref{fig:nonlocal_spin_Chern_data} (b). Note that the value at the origin ${\bf R=0}$ is the local Chern marker ${\cal C}^{s}({\bf r+0,r})={\cal C}^{s}({\bf r})$ that should remain unchanged within the same topological phase. Once again these features are exactly the same as their spinless counterparts in the Chern insulator, which have already been demonstrated previously\cite{Chen17,Chen19_AMS_review,Chen19_universality}.


\section{Conclusions}

In summary, we propose several optical absorption measurements for the ${\mathbb Z}_{2}$ invariant in 2D TR symmetric TIs. For the momentum space measurement, we first elaborate that the spin Berry curvature that integrates to the ${\mathbb Z}_{2}$ invariant is equivalently the Jacobian of the map between the $T^{D}$ BZ torus and the $S^{D}$ sphere of the Dirac Hamiltonian, thereby giving it a differential geometrical interpretation\cite{vonGersdorff21_unification}. Using a linear response theory\cite{Chen22_dressed_Berry_metric}, we then demonstrate that the spin Berry curvature is equivalently a spin-selected charge polarization susceptibility caused by circularly polarized electric field, which can be defined even at finite temperature and in the presence of many-body interactions. Our susceptibility formalism implies that the spin Berry curvature spectral function can be measured in pump-probe type of experiments using spin- and time-resolved ARPES. Moreover, performing the spin-resolved circular dichroism experiments merely at HSPs can already obtain the signs of the spin Berry curvature that are sufficient to determine the ${\mathbb Z}_{2}$ invariant. Using disordered BHZ model in a continuum as an example, we further demonstrate that despite protected by the band gap, strong enough perturbations can still affect the profile of spin Berry curvature.

For real space measurements, we first generalize the spin Chern marker to finite temperature via a linear response theory, yielding a dressed spin Chern marker that is equivalently a real space spin-selected charge polarization susceptibility caused by circularly polarized electric field. The nonlocal susceptibility corresponds to a spin Chern correlator, whose magnitude is much larger in the topologically trivial phase, and hence may be used to judge whether the system is moving towards a trivial or nontrivial phase upon tuning a parameter. A nonlocal spin Chern number is further defined from the off-diagonal elements of the spin Chern operator, which can be calculated from the Fourier transform of the spin Berry curvature, and consequently can be expressed as a spin-selected Wannier state correlation function. Due to the critical behavior of the spin Berry curvature, the nonlocal marker has a decay length that diverges at the TPT. As our linear response theory can take care of any many-body interactions in real experiments performed at finite temperature, we anticipate that these predictions that help to extract the ${\mathbb Z}_{2}$ invariant and the critical behavior can be readily tested in any 2D TR symmetric TIs.

\section{Acknowledgments}

The author acknowledges stimulating discussions with G. von Gersdorff, P. Molignini, B. Lapierre, and R. Chitra, as well as the financial support of the productivity in research fellowship from CNPq.

\vspace{1cm}

\bibliographystyle{unsrt}
\bibliography{Literatur}

\end{document}